# Differential Transmit Diversity Based on Quasi-Orthogonal Space-Time Block Code

Chau Yuen, Yong Liang Guan
School of Electrical and Electronic Engineering
Nanyang Technological University
Singapore
yuenchau@pmail.ntu.edu.sg, eylguan@ntu.edu.sg

Tjeng Thiang Tjhung
Institute for Infocomm Research
Singapore
tjhungtt@i2r.a-star.edu.sg

*Abstract*— By using joint modulation and customized constellation set, we show that Quasi-Orthogonal Space-Time Block Code (QO-STBC) can be used to form a new differential space-time modulation (DSTM) scheme to provide full transmit diversity with non-coherent detection. Our new scheme can provide higher code rate than existing DSTM schemes based on Orthogonal STBC. It also has a lower decoding complexity than the other DSTM schemes, such as those based on Group Codes, because it only requires a joint detection of two complex symbols. We derive the design criteria for the customized constellation set and use them to construct a constellation set that provides a wide range of spectral efficiency with full diversity and maximum coding gain.

*Keywords - differential space-time modulation; low decoding complexity; non coherent detection; quasi-orthogonal space-time block code.*

## I. INTRODUCTION

In wireless communications, system performance is often severely degraded by fading effects due to multi-path signal propagation. Modulation techniques designed for multiple transmit antennas, called space-time modulation or transmit diversity can be used to reduce fading effects effectively. Early transmit diversity schemes were designed for coherent detection, with channel estimates assumed available at the receiver. However, the complexity and cost of channel estimation grow with the number of transmit and receive antennas. Therefore the availability of transmit diversity schemes that do not require channel estimation is desirable. To this end, several differential space-time modulation (DSTM) schemes have been designed [1-8].

Hughes has designed a DSTM based on group codes [1, 2]; Hochwald and Sweldens have designed a DSTM based on unitary matrices [3]. Tarokh and Jafarkhani have proposed a DSTM by using the Orthogonal Space-Time Block Code (O-STBC) [4], while Ganesa and Stoica [8] provide another DSTM also based on O-STBC but with a much simpler decoding complexity. Hassibi et al. has proposed DSTM based on Cayley code and Sp(2). Recently, Al-Dhahir gives a new rate-two DSTM based on two parallel Alamouti O-STBCs for four transmit antennas system [6]. Among them the scheme in [8] has the simplest decoding complexity. However the DSTM in [8] is based on the O-STBC, whose maximum achievable code rate is limited to ¾ for four antennas and ½ for eight antennas when used with complex constellations.

In this paper, we propose a new DSTM scheme that is based on Quasi-Orthogonal STBC (QO-STBC) [9-13], to provide full transmit diversity with higher code rate than that of [8] and lower decoding complexity than other DSTM schemes [1-7]. We will derive the design criteria and construct an example constellation set for the proposed DSTM scheme. We will also study its decoding complexity and decoding performance in comparison with the existing schemes.

The organization of this paper is as follow, Section II reviews the DSTM signal model and the decoding performance criteria of DSTM. Section III proposes the new DSTM scheme based on QO-STBC. And Section IV gives the performance comparison. Finally Section V concludes the paper.

## II. REVIEW OF DSTM

### A. DSTM Signal Model

Consider a MIMO communication system, with $N_T$ transmit antennas and $N_R$ receive antennas. Let $\mathbf{H}_t$ be the $N_R \times N_T$ channel gain matrix at a time $t$. Thus the $ij^{th}$ element of $\mathbf{H}_t$ is the channel coefficient for the signal path from the $j^{th}$ transmit antenna to the $i^{th}$ receive antenna. Let $\mathbf{C}_t$ be the $N_T \times P$ codeword transmitted at a time $t$. Then, the received $N_R \times P$ signal matrix $\mathbf{R}_t$ can be written as

$$\mathbf{R}_t = \mathbf{H}_t \mathbf{C}_t + \mathbf{N}_t \quad (1)$$

where $\mathbf{N}_t$ is the additive white Gaussian noise. In this paper, the code length $P$ is set equal to $N_T$ as in [1] and [8], so that the transmitted codeword $\mathbf{C}_t$ is a square matrix.

At the start of the transmission, we transmit a known codeword $\mathbf{C}_0$, which is a unitary matrix of size $N_T \times N_T$. The codeword $\mathbf{C}_t$ transmitted at a time $t$ is differentially encoded by

$$\mathbf{C}_t = \mathbf{C}_{t-1} \mathbf{U}_t \quad (2)$$

where $\mathbf{U}_t$ is a unitary matrix (such that $\mathbf{U}_t \mathbf{U}_t^H = \mathbf{I}$) called the *code matrix*, that contains information of the transmitted data. Since $\mathbf{C}_0$ and $\mathbf{U}_t$ are both unitary, it follows that $\mathbf{C}_t$ is unitary for all time $t$. Hence the requirement for the code matrix $\mathbf{U}_t$ to be unitary is essential to ensure that all the transmitted codewords have constant power. If we assume that the channel remains unchanged during two consecutive code periods, i.e. $\mathbf{H}_t = \mathbf{H}_{t-1}$, then the received signal $\mathbf{R}_t$ at a time $t$ can be expressed [8] as

$$\mathbf{R}_t = \mathbf{R}_{t-1}\mathbf{U}_t + \tilde{\mathbf{N}}_t \qquad (3)$$

The received signal $\mathbf{R}_t$ can be differentially decoded as it depends only on the previous received signal block $\mathbf{R}_{t-1}$, the code matrix $\mathbf{U}_t$ and an equivalent additive white Gaussian noise $\tilde{\mathbf{N}}_t = -\mathbf{N}_{t-1}\mathbf{U}_t + \mathbf{N}_t$. Since $\mathbf{N}_t$ and $\mathbf{N}_{t-1}$ are white and $\mathbf{U}_t$ is unitary, it can be shown that the equivalent noise $\tilde{\mathbf{N}}_t$ is white, and has zero mean and twice the power as the channel noise $\mathbf{N}_t$ or $\mathbf{N}_{t-1}$ [7]. This explains why there is a 3dB SNR loss when the received signal is decoded differentially instead of coherently [1, 4, 7, 8]. The corresponding ML decoding decision metric is,

$$\begin{aligned}\hat{\mathbf{U}}_t &= \arg\min_{\mathbf{U}_t \in \mathcal{U}} \mathrm{tr}\left(\{\mathbf{R}_t - \mathbf{R}_{t-1}\mathbf{U}_t\}^H \{\mathbf{R}_t - \mathbf{R}_{t-1}\mathbf{U}_t\}\right) \\ &= \arg\max_{\mathbf{U}_t \in \mathcal{U}} \mathrm{Re}\left\{\mathrm{tr}\left(\mathbf{R}_t^H \mathbf{R}_{t-1}\mathbf{U}_t\right)\right\}\end{aligned} \qquad (4)$$

where $\mathcal{U}$ denotes the set of all possible code matrices.

*B. Diversity and Coding Gain*

The decoding performance of DSTM has been analyzed in [1], which found that the design criteria for DSTM are the same as for coherent space-time coding. Specifically, the transmit diversity level that can be achieved is given by:

$$\mathrm{Min}\left[\mathrm{rank}\left(\mathbf{U}_k - \mathbf{U}_l\right)\right] \qquad \forall k \neq l \qquad (5)$$

In order to achieve full transmit diversity, the minimum rank in (5) has to be equal to $N_T$.

For a full-rank DSTM code, its *coding gain* is defined in [1], [8] as

$$\mathrm{Min}\left[N_T \times \det\left((\mathbf{U}_k - \mathbf{U}_l)(\mathbf{U}_k - \mathbf{U}_l)^H\right)^{1/N_T}\right] \qquad \forall k \neq l \qquad (6)$$

In order to achieve optimum decoding performance, the coding gain has to be maximized.

III. NEW DSTM SCHEME

In this section, we shall develop a new DSTM scheme using the QO-STBC designed in [9] for four transmit antennas as an example. Our proposed technique, however, is applicable to any square QO-STBC that supports joint detection of two complex symbols, such as those in [10-13].

*A. Quasi Orthogonal Space-Time Block Code*

The $4 \times 4$ codeword of the QO-STBC in [9] (herein denoted as code Q4) is shown in (7), where $c_i$, $1 \leq i \leq 4$, represents the transmitted complex constellation symbol. It can be seen that the codeword of Q4, $\mathbf{C_{Q4}}$, is not a unitary matrix, since $\alpha \neq 1$ and $\beta \neq 0$ in general.

$$\mathbf{C_{Q4}} = \begin{bmatrix} c_1 & -c_2^* & -c_3^* & c_4 \\ c_2 & c_1^* & -c_4^* & -c_3 \\ c_3 & -c_4^* & c_1^* & -c_2 \\ c_4 & c_3^* & c_2^* & c_1 \end{bmatrix} \qquad (7)$$

$$\mathbf{C_{Q4}}\mathbf{C_{Q4}}^H = \begin{bmatrix} \alpha & 0 & 0 & \beta \\ 0 & \alpha & -\beta & 0 \\ 0 & -\beta & \alpha & 0 \\ \beta & 0 & 0 & \alpha \end{bmatrix} \neq \mathbf{I} \qquad (8)$$

where $\alpha = \sum_{i=1}^{4}|c_i|^2$; $\beta = 2 \times Re(c_1 \times c_4^* - c_2 \times c_3^*)$.

$\mathbf{C_{Q4}}$ can also be represented as [10]:

$$\mathbf{C_{Q4}} = \sum_{k=1}^{4}\left(\mathbf{A}_k c_k^R + j\mathbf{B}_k c_k^I\right) \qquad (9)$$

where $\mathbf{A}_k$ and $\mathbf{B}_k$ are called the dispersion matrices [14] with quasi-orthogonal algebraic properties described in [10], and the superscripts R and I represent the real and imaginary parts of a symbol respectively. The determinant of the codeword distance matrix of Q4 has been shown in [10] to be:

$$\det = \begin{bmatrix}(|\Delta_1 + \Delta_4|^2 + |\Delta_2 - \Delta_3|^2) \times \\ (|\Delta_1 - \Delta_4|^2 + |\Delta_2 + \Delta_3|^2)\end{bmatrix}^2 \qquad (10)$$

$$\Rightarrow \det_{\min} = \left[(|\Delta_1 + \Delta_4|^2) \times (|\Delta_1 - \Delta_4|^2)\right]^2 \text{ if } \Delta_2 = \Delta_3 = 0$$

where $\Delta_i$, $1 \leq i \leq 4$, represents the possible error in the $i^{\text{th}}$ transmitted constellation symbol. That is, $\Delta_i = c_i - e_i$ if the receiver decides erroneously in favor of $e_i$ if $c_i$ is transmitted. When considering the minimum determinant in accordance to (6), we can assume half of the codeword errors to be zero, which is similar to the design approach of constellation rotation for QO-STBC discussed in [10 – 13]. In order to achieve full transmit diversity and optimum coding gain in accordance with the design criteria defined in (5) and (6), the value in (10) has to be non-zero and maximized.

*B. New DSTM Based on QO-STBC*

As we have explained earlier, the use of a unitary code matrix, $\mathbf{U}_t$, is essential for formulating a DSTM scheme. In order to apply the code Q4 in a DSTM, i.e. use $\mathbf{C_{Q4}}$ as $\mathbf{U}_t$, $\mathbf{C_{Q4}}$ has to be a unitary matrix, i.e. $\alpha$ and $\beta$ in (8) have to be one and zero respectively for all possible values of $c_i$. This cannot be achieved if conventional memoryless modulation is applied. In order to have $\beta$ equal to zero, we can see from (8) that some form of correlation or constraint between the transmitted symbols $c_1$ and $c_4$, as well as between $c_2$ and $c_3$, are required. Therefore we propose the use of *joint modulation* with a *specially designed constellation set*, such that the transmitted symbol-pairs $\{c_1, c_4\}$ and $\{c_2, c_3\}$ can always achieve an $\alpha$ equal to one and a $\beta$ equal to zero.

Based on the above idea, a joint constellation set $\mathcal{M}$ which consists of $L$ sets of complex-valued constellation pair $\{x_k, y_k\}$ (where $x_k$ and $y_k$ are each a complex value, and $1 \leq k \leq L$) can provide a spectral efficiency of $R = 2(\log_2 L)/N_T$ bps/Hz. The requirements on the constellation set $\mathcal{M}$ are:

(i) $|x_k|^2 + |y_k|^2 = 0.5$

(ii) $\text{Re}(x_k y_k^*) = v$  (11)

(iii) maximize $\text{Min}\left\{\left[|\Delta x_{kl} + \Delta y_{kl}|^2 \times |\Delta x_{kl} - \Delta y_{kl}|^2\right]^2\right\}$

where $v$ can be any constant value, and $\Delta x_{kl} = x_k - x_l$, $\Delta y_{kl} = y_k - y_l$ for all $1 \leq k \neq l \leq L$. In complying with the Power Criterion (11)(i), we ensure that the transmitted codewords have a constant power and hence $\alpha = 1$. In complying with the Unitary Criterion (11)(ii), we ensure that $\mathbf{C}_{Q4}$ is always a unitary matrix, i.e. we achieve $\beta=0$. With the Performance Optimization Criterion (11)(iii), we ensure that the minimum determinant value in (10) is maximized such that the proposed DSTM can achieve full diversity and maximum coding gain.

To give an example of the proposed DSTM with joint modulation, consider the case of $R = 2$ bps/Hz (i.e. $L = 16$, $N_T = 4$). In the *encoding process*, we take four data bits and map it to one of the 16 constellation pairs $\{x_k, y_k\}$ in $\mathcal{M}$ as the code symbol pair $\{c_1, c_4\}$. Then we take the next four data bits and map them to another constellation pair in $\mathcal{M}$ as the code symbol pair $\{c_2, c_3\}$. Hence, the code matrix $\mathbf{U}_t$ in (2) can be obtained as $\mathbf{C}_{Q4}$ in (7) with the code symbols $c_1$ to $c_4$ assigned as above.

In the *decoding process*, with (9), the ML decision metric (4) can be simplified to:

$$\{\hat{c}_1, \hat{c}_4\} = \arg\max_{\{c_1, c_4\} \in \mathcal{M}} \left[ \begin{array}{l} \sum_{i=1,4} \text{Re}\{\text{tr}(\mathbf{R}_t^H \mathbf{R}_{t-1} \mathbf{A}_i)\} c_i^R + \\ \sum_{i=1,4} \text{Re}\{\text{tr}(\mathbf{R}_t^H \mathbf{R}_{t-1} j\mathbf{B}_i)\} c_i^I \end{array} \right]$$

$$\{\hat{c}_2, \hat{c}_3\} = \arg\max_{\{c_2, c_3\} \in \mathcal{M}} \left[ \begin{array}{l} \sum_{i=2,3} \text{Re}\{\text{tr}(\mathbf{R}_t^H \mathbf{R}_{t-1} \mathbf{A}_i)\} c_i^R + \\ \sum_{i=2,3} \text{Re}\{\text{tr}(\mathbf{R}_t^H \mathbf{R}_{t-1} j\mathbf{B}_i)\} c_i^I \end{array} \right]$$
(12)

As shown in (12), the ML decoding of the proposed differential QO-STBC modulation scheme can be achieved by the joint detection of two complex symbols, and the two ML metrics can be computed in parallel. So the proposed DSTM scheme does **not** increase the ML decoding complexity as compared to the original coherent Q4 code. Furthermore, it has a lower ML decoding complexity than the DSTMs reported in [1-7], which require generally a larger joint detection search space for the same spectral efficiency and antenna number (to be elaborated in Section IV).

### C. Design of Constellation Set

In this section, we propose a constellation set $\mathcal{M}$ that satisfies all three requirements in (11) with high flexibility in spectral efficiency. The proposed constellation set $\mathcal{M}$, $\{x_k, y_k\}$, $1 \leq k \leq L$, is:

$$\begin{cases} x_k = \dfrac{\exp[j(2k\pi/M)]}{\sqrt{2}} \\ y_k = 0 \end{cases} \quad \text{for } 1 \leq k \leq \dfrac{L}{2}$$

$$\begin{cases} x_k = 0 \\ y_k = \dfrac{\exp[j(2(k-L/2)\pi/M + \theta)]}{\sqrt{2}} \end{cases} \quad \text{for } \dfrac{L}{2} < k \leq L$$
(13)

where $M = L/2$ is an integer, and $\theta$ is a constellation rotation angle between 0 and $2\pi/M$.

Note that in (13), $x_k$ belongs to a half-power $M$-ary PSK constellation for $1 \leq k \leq L/2$, while $y_k$ belongs to a rotated half-power $M$-ary PSK constellation for $L/2 < k \leq L$. The parameter $M$ is related to the spectral efficiency $R$ by $R = 2(\log_2 M)/N_T$. Hence a DSTM with a wide range of spectral efficiency can be systematically designed from (13) by adjusting $M$. The constellation rotation angle $\theta$ provides an extra degree of freedom to maximize the diversity and coding gain of the resultant DSTM.

*Theorem 1*: For the DSTM constellation set defined in (13), the optimum value for the constellation rotation angle $\theta$ (in the sense of the Performance Optimization Criterion (11)(iii)) is $\pi/M$ when $M$ is even, and is $\pi/2M$ or $3\pi/2M$ when $M$ is odd.

*Proof of Theorem 1*: The proof is given in four cases, considering different values of $k$ and $l$.

<u>Case 1: $1 \leq k, l \leq L/2$, and $k \neq l$.</u>

In this case, $\Delta y_{kl}$ is always zero (since $y_k$ and $y_l$ are always zero), but not $\Delta x_{kl}$ (since $k \neq l$), hence the determinant value in (11)(iii) can never be zero, and its value is independent of $\theta$. Hence in this case, full diversity is always achieved, and the coding gain optimization does not depend on $\theta$.

<u>Case 2: $L/2 < k, l \leq L$ and $k \neq l$.</u>

The proof is similar to Case 1 with the role of $\Delta x_{kl}$ and $\Delta y_{kl}$ interchanged.

<u>Case 3: $1 \leq k \leq L/2$ and $L/2 < l \leq L$.</u>

For this case, the determinant value in (11)(iii) can be simplified as follows:

$$\det = \frac{1}{16}\left[\left(\left|\exp\left[j\frac{2k\pi}{M}\right] + \exp\left[j\left(\frac{2(l-L/2)\pi}{M}+\theta\right)\right]\right|^2\right) \times \left(\left|\exp\left[j\frac{2k\pi}{M}\right] - \exp\left[j\left(\frac{2(l-L/2)\pi}{M}+\theta\right)\right]\right|^2\right)\right]^2$$

$$= \left(\sin^2\left(\frac{2\pi k}{M} - \frac{2\pi m}{M} - \theta\right)\right)^2 \quad 1 \le k, m \triangleq l - \frac{L}{2} \le M \quad (14)$$

$$= \sin^4\left(\frac{2(n-p)\pi}{M}\right) \quad 0 \le n \triangleq k - m \le M - 1$$

where $k$, $l$, $m$, $n$ are integers, and $M \triangleq L/2$ $p \triangleq M\theta/2\pi$ is a real number between 0 and 1 inclusively.

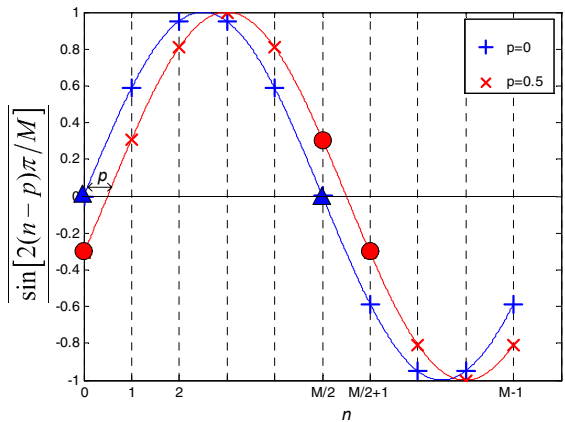

(a) $M$ even

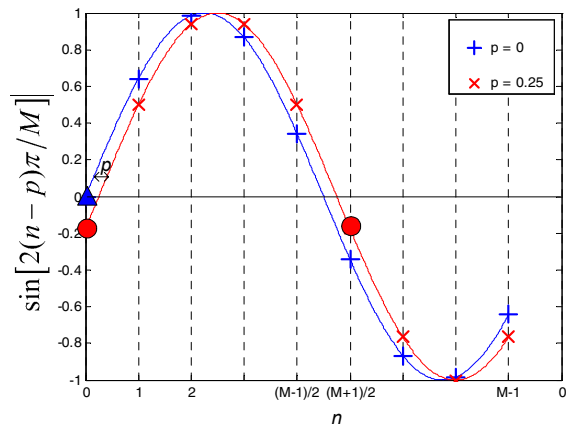

(b) $M$ odd

Figure 1. Optimization of constellation rotation angle $\theta = 2\pi p/M$

To maximize the determinant value in (14), we first consider even values of $M$. As shown in Figure 1(a), the solid line, $\sin[2n\pi/M]$, is zero if $n = 0$ or $M/2$, as shown by the *triangular* markers. In order to achieve full diversity, a right shift, $p$, can be introduced to obtain the dash line, $\sin[2(n-p)\pi/M]$, such that its value is non-zero for all integer values of $n$. To further achieve maximum coding gain, the value of $p$ should be chosen such that the minimum *absolute* values of $\sin[2(n-p)\pi/M]$ at integer values of $n$ are maximized, as shown by the *circular* markers in Figure 1(a). This corresponds to $p = 0.5$ and $\theta = \pi/M$. Similarly, for odd values of $M$, the optimum $p$ value is 0.25 or 0.75, which corresponds to $\theta = \pi/2M$ or $3\pi/2M$.

*Case 4: $1 \le l \le L/2$ and $L/2 < k \le L$.*

The proof is similar to Case 3. As the determinant value in (11)(iii) does not depend on $\theta$ for Cases 1 and 2, and the optimum $\theta$ value has been derived for Cases 3 and 4, *Theorem 1* is proved. ∎

IV. PERFORMANCE RESULTS

The coding gain and decoding complexity of our proposed DSTM with an optimized constellation set and some existing DSTM [1, 2, 8] are shown in Table I. It can be seen that our proposed DSTM provides the highest coding gain when compared with DSTM based on O-STBC [8] and DSTM based on group codes [1,2]. Our proposed DSTM also has a lower decoding complexity than those in [1, 2], because the ML decoder of our proposed DSTM only needs to jointly decode two complex symbols in parallel. At 1.5 (or 2) bps/Hz, this leads to a ML search space of 8 (or 16) with 4PSK (or 8PSK) constellation, compared with a search space of 64 (or 256) for the schemes in [1, 2] with 64PSK (or 256PSK) constellation. Although the DSTM based on O-STBC in [8] has the lowest decoding complexity, it has a lower coding gain, and it does not has the flexibility to provide 2 bps/Hz using conventional constellations (hence it is not included in the comparison table).

TABLE I. COMPARISON OF CODING GAINS AND DECODING COMPLEXITY FOR DSTM FOR FOUR TRANSMIT ANTENNAS

| DSTM | Constellation | Bps/Hz | Coding Gain | ML Search Space |
|---|---|---|---|---|
| [1, 2] | 64PSK | 1.5 | 1.85 | 64 |
| Proposed scheme | (13) with $M = 4$, $\theta = \pi/4$ | 1.5 | 2.83 | 8 |
| [8] | QPSK | 1.5 | 2.70 | 4 |
| [1, 2] | 256PSK | 2 | 0.78 | 256 |
| Proposed scheme | (13) with $M = 8$, $\theta = \pi/8$ | 2 | 1.17 | 16 |

In Figure 2, we compare the block error rate (BLER) performance of our proposed DSTM with those of the other DSTM's reported in the literature [5, 7, 8]. Comparing our proposed DSTM of 2 bps/Hz with Cayley's DSTM with spectral efficiency of 1.75 bps/Hz (obtained with $Q=7$, $r=2$ in [5]), our DSTM provides a comparable BLER performance but at a higher spectral efficiency. Compared with the DSTM based on Sp(2) with spectral efficiency of 1.94 bps/Hz (obtained with $M=5$, $N=3$ in [7]) and ML search space of 225, our proposed DSTM performs 1dB worse in BLER but has higher spectral efficiency and a much smaller ML search space of 16. Finally, the DSTM based on O-STBC [8] has the highest spectral efficiency of 2.25 bps/Hz, but it has the worst BLER performance.

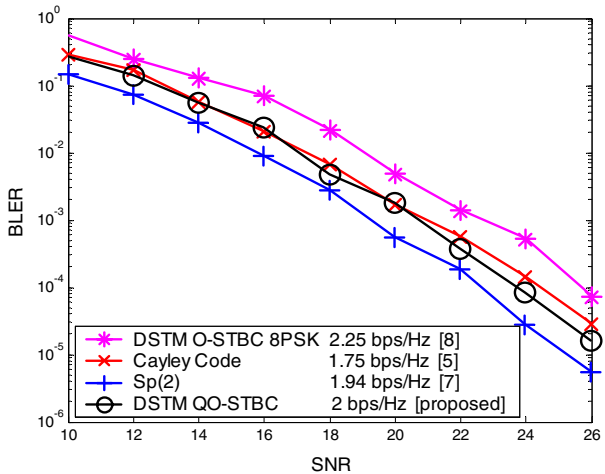

Figure 2. Simulated block error rate (BLER) of DSTM for four transmit and one receive antennas

In Figure 3, we compare the bit error rate (BER) performance of our proposed DSTM with differential and coherent detection, with that of Q4 with constellation rotation [10-13] and that of the rate-two DSTM scheme of [6]. All codes considered in this figure have the same spectral efficiency of 2 bps/Hz.

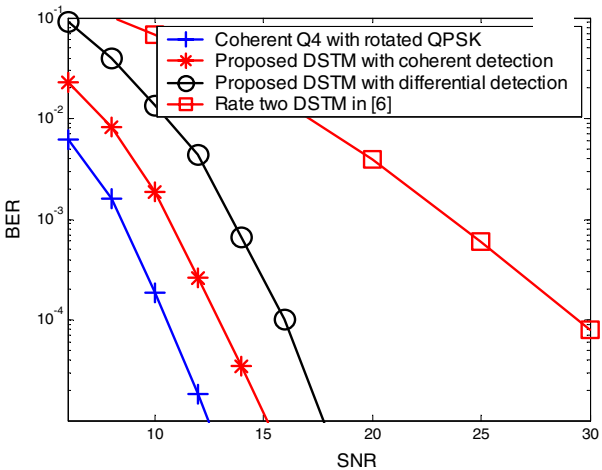

Figure 3. Simulated bit error rate (BER) of DSTM schemes for four transmit and two receive antennas with spectral efficiency of 2 bps/Hz

Figure 3 shows that there is a 3dB SNR gap between the coherent and differential detection of our proposed DSTM. This is due to the doubling of noise power for the latter, as explained earlier in this paper. Compared to the original QO-STBC, Q4 with constellation rotation and coherent detection [10-13], our proposed DSTM is 2dB worse if coherently detected. This loss is due to the constraint that the codeword in our proposed scheme must be a unitary matrix in order to support differential modulation. However, this loss is acceptable, as a coherent QO-STBC will suffer performance degradation if the channel estimation is imperfect. Finally, Figure 3 also shows that our proposed scheme has a better BER performance than the DSTM presented in [6], which does not achieve full diversity.

## V. CONCLUSIONS

We propose a new differential space-time modulation (DSTM) scheme based on QO-STBC. The main idea is to force the QO-STBC codeword to be a unitary matrix by using joint symbol modulation with a specially designed constellation set. The design criteria for the corresponding constellation set to achieve full diversity and maximum coding gain are derived. We then propose a constellation set (comprising zero, PSK and rotated PSK symbols) that meets the design criteria and supports a wide range of spectral efficiency. Our proposed DSTM scheme has the merits of QO-STBC, hence it achieves higher code rate than DSTM based on O-STBC, and lower ML decoding complexity (smaller search space) than other DSTMs, such as $[1-7]$, with a comparable decoding performance. Although a four-antenna code is considered in this letter, the differential code design and code optimization ideas proposed herein apply to any QO-STBCs that can be found in the literature, including those for eight transmit antennas.

## REFERENCES

[1]  Hughes, B. L., "Differential space-time modulation", *IEEE Trans. on Info. Theory*, Vol:46, Issue:7, Nov. 2000, 2567–2578.

[2]  Hughes, B. L., "Optimal space-time constellations from groups", *IEEE Trans. on Info. Theory*, Vol:49, Issue:2, Feb. 2003, 401– 410.

[3]  Hochwald, B. M.; Swelders, W., "Differential unitary space-time modulation", *IEEE Trans. on Comms.*, Vol:48, Issue:12, Dec. 2000, 2041–2052.

[4]  Tarokh, V.; Jafarkhani, H., "A differential detection scheme for transmit diversity", *IEEE Journal on Selected Areas in Comms.*, Vol:18, Issue:7 , July 2000, 1169–1174.

[5]  Hassibi, B.; Hochwald, B. M., "Cayley differential unitary space-time codes", *IEEE Trans. On Info. Theory*, 2002, 1458 – 1503.

[6]  Al-Dhahir, N., "A new high-rate differential space-time block coding scheme", *IEEE Comms. Letters*, Vol:7, Issue:11, Nov. 2003, 540–542.

[7]  Jing, Y.; Hassibi, B., "Design of fully-diverse multi-antenna codes based on Sp(2)", *ICASSP 2003*, Vol: 4, 33-36.

[8]  Ganesan, G.; Stoica, P., "Differential space-time modulation using space-time block codes", *IEEE Signal Processing Letters*, Vol:9 Issue:2 , Feb 2002, 57 –60.

[9]  Jafarkhani, H., "A quasi-orthogonal space-time block code", *IEEE Trans. on Comms.,* Vol:49 Issue:1, Jan. 2001, 1-4.

[10]  Yuen, C.; Guan, Y. L.; Tjhung, T. T. , "Full-Rate Full-Diversity STBC with Constellation Rotation". *VTC 2003-Spring,* Vol:1, 296 –300.

[11]  Tirkkonen, O., "Optimizing STBC by Constellation Rotations", *FWCW 2001*, 59-60.

[12]  Sharma, N.; Papadias, C.B., "Improved quasi-orthogonal codes through constellation rotation", *IEEE Trans. on Comms.*, Vol:51, Issue:3, March 2003, 332- 335.

[13]  Su, W.; Xia, X., "Quasi-Orthogonal STBC with Full Diversity", *Globecom 2002*, 1098 –1102.

[14]  Hassibi, B.; Hochwald, B. M., "High-Rate Codes that are Linear in Space and Time", *IEEE Trans. on Info. Theory*, Vol:48 Issue:7 , Jul 2002, 1804 –1824.